\documentclass[conference]{IEEEtran} 
\usepackage[ruled,linesnumbered]{algorithm2e}

\IEEEoverridecommandlockouts

\usepackage{cite}
\usepackage{amsmath, amssymb, amsfonts, mathtools}
\usepackage{algorithmic}
\usepackage{graphicx}
\usepackage{textcomp}
\usepackage{xcolor}
\def\BibTeX{{\rm B\kern-.05em{\sc i\kern-.025em b}\kern-.08em
    T\kern-.1667em\lower.7ex\hbox{E}\kern-.125emX}}

\graphicspath{{./Figures/}}
\newtheorem{Remark}{Remark}

\newtheorem{Lemma}{Lemma}
\newtheorem{Theorem}{Theorem}
\newtheorem{Corollary}{Corollary}

\DeclareMathAlphabet{\mathit}{OT1}{bch}{m}{it}

\DeclareFontFamily{U}{mathx}{}
\DeclareFontShape{U}{mathx}{m}{n}{<-> mathx10}{}
\DeclareSymbolFont{mathx}{U}{mathx}{m}{n}
\DeclareMathAccent{\widehat}{0}{mathx}{"70}
\DeclareMathAccent{\widecheck}{0}{mathx}{"71}

\newcommand{\mean}{\mathbf{E}}
\newcommand{\mgf}{\mathbf{M}}

\newcommand{\eff}{\textnormal{eff}}

\newcommand{\tsum}{\mathop{\scalebox{0.8}{$\displaystyle\sum$}}}

\newcommand{\snr}{{\gamma}}
\newcommand{\avgsnr}{\bar{\gamma}}
\newcommand{\los}{{\textnormal{LoS}}}

\begin{document}

\title{Fluctuating Line-of-Sight Fading Distribution: Statistical Characterization and Applications \\
}

\author{\IEEEauthorblockN{
        Thanh~Luan~Nguyen\IEEEauthorrefmark{1}, 
        Georges~Kaddoum{\IEEEauthorrefmark{1}$^{,}$\IEEEauthorrefmark{2}}
        }
    \IEEEauthorrefmark{1} Department of Electrical Engineering, \'{E}cole de Technologie Sup\'{e}rieure (\'{E}TS), Montr\'{e}al, QC, Canada \\
    \IEEEauthorrefmark{2} Artificial Intelligence \& Cyber Systems Research Center, Lebanese American University
    \\
      Emails: 
        \IEEEauthorrefmark{1}thanh-luan.nguyen.1@ens.etsmtl.ca,
        \IEEEauthorrefmark{1}$^{,}$\IEEEauthorrefmark{2}georges.kaddoum@etsmtl.ca, 
}

\maketitle

\begin{abstract}
We introduce the fluctuating Line-of-Sight (fLoS) fading model, characterized by parameters $K$, $k$, $\lambda$, and $\Omega$. 
    The fLoS fading distribution is expressed in terms of the multivariate confluent hypergeometric functions $\Psi_2$, $\Phi_3^{(n)}$, and $\Phi_3 = \Phi_3^{(2)}$ and encompasses well-known distributions, such as the Nakagami-$m$, Hoyt, Rice, and Rician shadowed fading distributions as special cases. An efficient method to numerically compute the fLoS fading distribution is also addressed.
Notably, for a positive integer $k$, the fLoS fading distribution simplifies to a finite mixture of $\kappa$-$\mu$ distributions.
    Additionally, we analyze the outage probability and Ergodic capacity, presenting a tailored Prony's approximation method for the latter. 
Numerical results are presented to show the impact of the fading parameters and verify the accuracy of the proposed approximation. Moreover, we illustrate an application of the proposed fLoS fading distribution for characterizing wireless systems affected by channel aging.
\end{abstract}

\begin{IEEEkeywords}
$\kappa$-$\mu$ fading, confluent hypergeometric functions, outage probability, channel capacity, Prony's method
\end{IEEEkeywords}

\section{Introduction}


\IEEEPARstart{T}{he} advent of sixth-generation (6G) wireless networks is expected to bring forth various applications for ubiquitous, high-speed connectivity \cite{Jiang2021OJCS, Wang2023CST}. 
    To name a few, the use reconfigurable intelligent surfaces (RIS) can configure radio environments for enhanced performance. Moreover, high-mobility communication (HMC) ensures broadband services for users traveling at speeds of up to $1000$ km/h. 
In addition, the utilization of terahertz (THz) bands, ranging from $0.1$ THz to $10$ THz, offer transmission rates reaching multiple terabits per second. 
    However, despite these advancements, significant challenges persist within each of these technologies. 
For instance, RIS-aided communication requires compelling applications that demonstrate substantial performance gains over existing technologies like mmWave and massive multiple-input-multiple-output (mMIMO) systems \cite{Bjornson2020CMag}.
    HMC faces issues like channel aging, where the system relies on outdated channel state information (CSI) for information processing \cite{ChopraTWC2018}. 
THz communications encounter blockages and lack convincing theoretical frameworks and practical channel models \cite{GeraciCM2023, AzariCST2022}. 
    Therefore, to fully realize the potential of these emerging technologies and address the associated challenges, the integration of novel fading channel models that accurately simulate real-world 6G environments is essential. 

The literature presents numerous generalized fading models that effectively depict the statistical behavior of mobile radio signals. 
    Various distributions, including the Rayleigh, Rice, Nakagami-$m$, Hoyt (Nakagami-$q$), and $\kappa$-$\mu$ distributions, describe short-term signal fluctuations \cite{Wang2020VTM}. 
Long-term variations, usually referred to as shadowing, are represented by the lognormal distribution, assuming independent attenuation from obstacles \cite{Wang2020VTM, Gutierrez2022TVT}. 
    However, practical scenarios introduce dependencies among obstacle attenuations due to phenomena like diffraction and refraction, challenging the independence assumption. 
While the lognormal model is widely accepted, alternative approaches like the ${\text{Nakagami-}m}$ distribution provide tractable substitute \cite{Dang2023AWPL}. 
    Moreover, in the context of dynamic environments, such as in vehicle-to-vehicle (V2V) communications, Gaussian, Rayleigh, and Rician distributions are favored for their ability to account for non-Gaussian shadowing with non-zero skewness in the dB scale \cite{Wang2020ECAP, Gutierrez2022TVT}. 
Despite these models, significant theoretical gaps persist in characterizing generalized shadowing and other Line-of-Sight (LoS) variations due to various Radio Frequency (RF) propagation phenomena, leading to the need of comprehensive models for capturing real-world signal propagation phenomena.

This paper seeks to address these gaps by providing a novel approach for characterizing the fluctuations of the dominant specular component, i.e., the LoS component, using generalized distributions. Specifically, we propose a novel fluctuating LoS (fLoS) fading distribution, where LoS fluctuations are modeled by a noncentral chi-square distribution. This fLoS fading distribution encompasses classical models, like the Hoyt, Rice, Nakagami-$m$, Rician shadowed, and mixture of $\kappa$-$\mu$ distributions, as special cases.
    Our main contributions are summarized as follows\footnote{Our results are reproducible with MATLAB code, uploaded at https://github.com/thanhluannguyen/fluctuatingLoS}:
\begin{itemize}
    \item We introduce a novel probability density function (PDF) and cumulative distribution (CDF) for the proposed fLoS fading characterized by $4$ parameters $K$, $k$, $\lambda$, and $\Omega$.
    \item We offer numerical methods for tackling the derived distributions due to their high computational complexity. 
        Notably, for integer $k$, the fLoS fading distribution can be expressed as a mixture of $\kappa$-$\mu$ distributions.
    \item We derive the outage probability (OP) and Ergodic capacity (EC) by introducing a Prony's method to approximate various functions and compute the EC for wireless systems.
    \item We present a use case for the fLoS fading distribution and we analyze the wireless system under the influence of channel aging. 
\end{itemize}


The paper is organized as follows: Section \ref{sec:II} introduces the physical model of the fLoS fading channel. Subsequent analysis of the fLoS fading distribution, including the PDF, CDF, moments, and moment generating function (MGF), is presented in Section \ref{sec:statCharac}. Section \ref{sec:performAna_App} presents the performance analysis and a potential application for the fLoS fading distribution. Section \ref{sec:result} details the numerical results. The paper is concluded with a summary of findings in Section \ref{sec:conclusion}.

\section{Physical Model}
\label{sec:II}

We consider the first-order scattering model, introduced in \cite{lopez2022TVT, Salo2006TAP}, to present 
the received signal under the fLoS fading model with LoS fluctuations as follows:
\begin{align}
    S = \omega_0 \xi e^{j\phi_0} + \sigma G, \label{eq:S_1}
\end{align}
where $\omega_0 e^{j\phi_0}$ is the LoS (specular) component with constant amplitude $\omega_0$, $\phi_0$ is a random variable (RV) uniformly distributed in $[0, 2\pi)$. The term $\sigma G$, where $G \sim {\cal CN}(0,1)$, represents the diffuse scattering, or non-LoS component, with total power $\sigma^2$. 
    Here, ${K = \frac{|\omega_0|^2}{\sigma^2}} \in [0, \infty)$, such that ${\omega_0 = \frac{\sqrt{K}}{\sqrt{K+1}}}$ and ${\sigma = \frac{1}{\sqrt{K+1}}}$, is the ratio between the power of the dominant LoS component and the average power of the scattering components.
In various researches, the component $\xi$ is often assumed as the lognormal shadowing, or normal (Gaussian) shadowing in decibels. 
    This model assumes independent attenuations from independent obstacles, i.e., scatterers and reflectors, within local clusters, common in heavy shadowing environments. 
Specifically, if each obstacle (reflector or scatterer) independently attenuates the signal by $A_m$ along a path, the total attenuation due to $M$ independent obstacles is $\xi = \prod_m^M A_m$ or $\xi|{\text{dB}} = \sum_m^M \xi_m|{\text{dB}}$ in dB. With sufficiently large $M$, the total attenuation (in dB) follows a Gaussian distribution due to the Central Limit Theorem (CLT) regardless of the distribution of $A_m$ distribution \cite{Gutierrez2022TVT}. 
    Yet, in reality, the obstacles' attenuations are often correlated due diffraction and receiver body vibrations. In high frequency bands (e.g., mmWave, THz), molecular absorption loss significantly influences obstacle attenuation \cite{Wang2023CST}. Consequently, the characterization of the shadowing distribution is useful when $M$ is sufficiently small to prevent the probability of $\xi=0$ approaching one. However, when $M$ is small, the CLT cannot be applied to characterize $\xi$ with a lognormal distribution. Despite the widely accepted lognormal assumption, there exist other tractable shadowing models \cite{Dang2023AWPL}. 
    In this manner, we consider $\xi^2$ follows a (scaled) noncentral chi-squared distribution with $2k$ degrees of freedom (d.o.f.), $2\lambda$ noncentrality parameter, and scale factor $\frac{\Omega}{2}$, denoted as $\xi^2 \sim \widetilde{\chi}_k^2(\Omega; \lambda)$.
The PDF of $\xi^2$ is:
\begin{align}
\label{eq:ncx2_pdf}
f_\xi(\xi) = \frac{e^{-\frac{x}{\Omega}-\lambda}}{\Omega} 
    \bigg(
        \frac{x}{\lambda \Omega}
    \bigg)^{\frac{k-1}{2}}
    I_{k-1}\bigg(
        2 \sqrt{\frac{\lambda x}{\Omega}}
    \bigg), x > 0.
\end{align}
where $I_v(x)$ is the $v$th order modified Bessel function of the first kind. 
It is noted that for $\Omega = \frac{1}{k+\lambda}$ with $\lambda \to 0$, \eqref{eq:ncx2_pdf} reduces to the Gamma distribution with unit mean and shape factor $k$,~as $f_{\xi}(x) = \frac{k^k}{\Gamma(k)} x^{k-1} e^{-k x}$, $x > 0$, and the model \eqref{eq:S_1} is simplified to the well-known Rician shadowed fading, where $k$ depicts the shadowing severity \cite{Abdi2003TWC}.
    We must note here that while the term "shadowing" is used, the fLoS model also includes variations in the direct LoS path due to other factors, such as movement or dynamic environments.

\begin{Remark}
The noncentral chi-square distribution is closely related to the $\kappa$-$\mu$ fading power distribution with parameters $(\avgsnr; \kappa, \mu)$. 
    These two distributions can be deduced from one another by substituting suitable parameters. 
Specifically, let ${\cal F}_{\kappa\mu}(\bar{\gamma}; \kappa, \mu)$ be a RV following the $\kappa$-$\mu$ fading power distribution, we have the following relations:
\begin{align}
\widetilde{\chi}_{k}^2(\Omega; \lambda)
    &\mathop{=}^d {\cal F}_{\kappa\mu}(\Omega(k+\lambda); \lambda / k, k) , \\
{\cal F}_{\kappa\mu}(\avgsnr; \kappa, \mu)
    &\mathop{=}^d \widetilde{\chi}_{\mu}^2(\bar{\gamma} [\mu (1+\kappa)]^{-1}; \mu\kappa).
\end{align}
%
%
\end{Remark}

\section{Statistical Characterization}
\label{sec:statCharac}

In this section,  we derive exact analytical formulas for the Moment Generating Function (MGF), the PDF, the CDF, and raw moments of the instantaneous SNR under fLoS fading conditions.
    For normalization purposes, let $\Omega = \frac{1}{k+\lambda}$, so that $\mean\{|S|^2\} = 1$ and $\snr = \avgsnr |S|^2$, where $\avgsnr$ is the average~SNR.

\subsection{Probability Functions}

\begin{Lemma}
\label{lem:mgf_snr}
Considering $\snr \sim {\cal F}_{\los}(\avgsnr; K; k, \lambda, \Omega)$ with non-negative real values $K$, $k$, $\lambda$, and $\Omega$, to characterize the SNR under fLoS fading, where $\avgsnr = \mean\{ \snr\}$ denotes the average SNR. Then, its MGF is obtained as:
\begin{align}
{\bf M}_{\snr}(s) 
    &=  \frac{1}{1 - \sigma^2 \avgsnr s}
    {\bf M}_{\xi^2}\left( 
        \frac{|\omega_0|^2 \avgsnr s}{1- \sigma^2 \avgsnr s}
    \right) \label{eq:mgf_gamma_1} \\
    &=  [1-\sigma^2 \avgsnr s]^{k-1}
    \frac{\exp\left(
        \frac{\lambda \Omega |\omega_0|^2 \avgsnr s }{1-(\sigma^2+\Omega |\omega_0|^2) \avgsnr s}
    \right)}{[1-(\sigma^2+\Omega |\omega_0|^2) \avgsnr s]^k}
    , \label{eq:mgf_gamma_2}
\end{align}
for $\avgsnr s < (\sigma^2 +\Omega |\omega_0|^2)^{-1}$.
\end{Lemma}
\begin{IEEEproof}
    We can characterize the SNR based on \eqref{eq:S_2} as:
\begin{align}
\snr|\xi 
    &\sim \avgsnr \left| {\cal CN}\left( \omega_0 \xi e^{j\phi_0}, \sigma^2 \right) \right|^2 
\label{eq:S_2} \\
    &\sim \frac{\sigma^2 \avgsnr}{2}
    \bigg[
            {\cal N}^2\left( \sqrt{2} \omega_0 \sigma^{-1} \xi \cos\phi_0, 1 \right)
\nonumber\\
    &\qquad\qquad
        + {\cal N}^2\left( \sqrt{2} \omega_0 \sigma^{-1} \xi \sin\phi_0, 1 \right)
    \bigg] \\
    &\mathop{\sim}\limits^{(a)} 
    \widetilde{\chi}_{1}^2\left(
        \sigma^2 \avgsnr, |\omega_0|^2 \sigma^{-2} \xi^2
    \right),
\end{align}
where $(a)$ is due to the fact that the sum of two statistically independent squared normal (Gaussian) RVs with unit variances follows the noncentral chi-square distribution. 
Hence, conditioned on $\xi$, the SNR follows the noncentral chi-square distribution with $2$ d.o.f., noncentrality parameter $\frac{2 |\omega_0|^2}{\sigma^2} \xi^2$, and scale factor $\frac{\sigma^2}{2}$.
    The conditional MGF of $\snr$ on $\xi$ is:
\begin{align}
\mathbf{M}_{\snr|\xi}(s|\xi) = \frac{1}{1 - \sigma^2 \avgsnr s} 
\exp\left( {\frac{|\omega_0|^2 \avgsnr s}{1- \sigma^2 \avgsnr s} \xi^2} \right),
\end{align}
for $\avgsnr \sigma^2 s < 1$. The marginal MGF of the SNR is obtained by taking the expectation of $\mathbf{M}_{\snr|\xi}(s|\xi)$ over $\xi$, which results in \eqref{eq:mgf_gamma_1}. 
    Finally, applying $\mgf_{\xi^2}(s) = \frac{1}{(1-\Omega s)^k} e^{\frac{\Omega \lambda s}{1-\Omega s}}$ yields \eqref{eq:mgf_gamma_2}, which completes the proof of Lemma \ref{lem:mgf_snr}.
\end{IEEEproof}

\begin{Lemma}
Considering $\snr \sim {\cal F}_{\los}(\avgsnr; K; k, \lambda, \Omega)$ with non-negative real values $K$, $k$, $\lambda$, and $\Omega$, where $\avgsnr = \mean\{ \snr\}$ denotes the average SNR. Then, the $n$th moment of $\snr$ is
\begin{align}
\mean\left\{ \snr^n \right\}
    =   n! (\sigma^2 \avgsnr)^n \sum_{i=0}^{n} \binom{n}{i} \left( \frac{\Omega |\omega_0|^2}{\sigma^2} \right)^{i}
    L_i^{k-1}(-\lambda),
\label{eq:raw_moment}
\end{align}
where $L_n^\alpha(x)$ denotes the generalized Laguerre polynomial.
\end{Lemma}

\begin{Theorem}
\label{theo:pdf_snr}
Considering $\snr \sim {\cal F}_{\los}(\avgsnr; K; k, \lambda, \Omega)$ with non-negative real values $K$, $k$, $\lambda$, and $\Omega$, where $\avgsnr = \mean\{ \snr\}$ denotes the average SNR. Then, the PDF of $\snr$ is:
\begin{align}
f_{\snr}(x) &=
    \frac{1}{\sigma^2 \avgsnr [1+\Omega|\omega_0|^2 \sigma^{-2}]^k }
    \exp\left(
        -   \frac{x}{\sigma^2 \avgsnr} - \lambda
    \right)
    \nonumber\\
    &\quad\times
    \widetilde{\Psi}_2\left(
        k; \frac{\lambda}{1+\frac{\Omega |\omega_0|^2}{\sigma^2}}, \frac{\Omega |\omega_0|^2}{\sigma^2+\Omega |\omega_0|^2} \frac{x}{\sigma^2 \avgsnr}
    \right),
\label{eq:pdf_gamma}
\end{align}
for $x > 0$, where $\widetilde{\Psi}_2(a; w, z) \triangleq \Psi_2(a; 1, a; z, w)$ with $\Psi_2(\cdot)$ being the confluent hypergeometric function of two variables \cite[Eq. (1.4.9)]{srivastava1985multiple}.
\end{Theorem}

\begin{IEEEproof}
The PDF of $\snr$ is obtained by taking the inverse Laplace transform of ${\bf M}_{\snr}(-s)$ from to $x$-domain as:
\begin{align}
f_{\snr}(x) 
    &= \mathbf{L}^{-1}\left\{ {\bf M}_{\snr}(-s); s, x \right\} \\
    &= \exp\left( -\frac{x \avgsnr^{-1}}{\sigma^2 + \Omega|\omega_0|^2} \right)
\nonumber\\
    &\qquad\times
        \mathbf{L}^{-1}\!\left\{ {\bf M}_{\snr}\left( -s + \frac{\avgsnr^{-1}}{\sigma^2 + \Omega|\omega_0|^2} \right); s, x \right\},
\end{align}
where the last equality is due to the frequency shifting property of the Laplace transform. After that, utilizing \cite[Eq. (7)]{Ermolova2014TSC} and \cite[Eq. (16)]{Ermolova2014TSC}, we get:
\begin{align}
    \mathbf{L}^{-1}\left\{
        \frac{(a+s)^{k-1}}{s^k} e^{\frac{a b}{s}}; s, x
    \right\} &= \Phi_3(1-k; 1; -a x, a b x) \\
        &= e^{-ax-b} \widetilde{\Psi}_2(k; b, a x),
\end{align}
and after some mathematical manipulations, we obtain \eqref{eq:pdf_gamma}. This completes the proof of Theorem \ref{theo:pdf_snr}.
\end{IEEEproof}

\begin{Theorem}
\label{theo:theo_cdf}
Considering $\snr \sim {\cal F}_{\los}(\avgsnr; K; k, \lambda, \Omega)$ with non-negative real values $K$, $k$, $\lambda$, and $\Omega$, where $\avgsnr = \mean\{ \snr\}$ denotes the average SNR. Then, the CDF of $\snr$ is:
%
%
\begin{align}
F_{\snr}(x) &=
    \frac{x}{\sigma^2 \avgsnr}
    A^k
    \exp\left(
        -   \frac{A}{\sigma^2 \avgsnr} x - B\lambda
    \right)
    \nonumber\\
    &\quad\times
    \Phi_3^{(3)}\bigg(
        1, 1-k; 2; \frac{A}{\sigma^2 \avgsnr} x,
        - \frac{B}{\sigma^2 \avgsnr}x, 
        \frac{A B \lambda}{\sigma^2 \avgsnr} x
    \bigg),
\label{eq:cdf_snr}
\end{align}
for $x > 0$, where $A \triangleq \frac{\sigma^2}{\sigma^2+\Omega |\omega_0|^2}$, $B \triangleq \frac{\Omega |\omega_0|^2}{\sigma^2+\Omega |\omega_0|^2}$, and~$\Phi_3^{(n)}(\cdot)$ is the $n$-variate confluent hypergeometric function \cite[Eq. (1.4.17)]{srivastava1985multiple}.
\end{Theorem}

\begin{IEEEproof}
The CDF of $\snr$ is obtained by taking the inverse Laplace transform of $s^{-1} {\bf M}_{\snr}(-s)$ as:
\begin{align}
F_{\snr}(x) 
    &= \mathbf{L}^{-1}\left\{ s^{-1} {\bf M}_{\snr}(-s); s, x \right\} \\
    &= \exp\left( -\frac{x \avgsnr^{-1}}{\sigma^2 + \Omega|\omega_0|^2} \right)
\nonumber\\
    &\qquad\times
        \mathbf{L}^{-1}\!\left\{ \frac{1}{s} \frac{{\bf M}_{\snr}\left( -s + \frac{\avgsnr^{-1}}{\sigma^2 + \Omega|\omega_0|^2} \right)}{1-\frac{\avgsnr^{-1}}{\sigma^2 + \Omega|\omega_0|^2} \frac{1}{s} }; s, x \right\}.
\end{align}

Note that $\Phi_3^{(n)}(\cdot)$ is expressed as \cite[Eq. (1.4.17)]{srivastava1985multiple}:
\begin{align}
&\Phi_3^{(n)}(b_1, \dots, b_{n-1}; c; d_1 t, \dots, d_n t)
\nonumber\\
&\qquad
=   \lim_{b_n \to \infty}
    \Phi_2^{(n)}(b_1, \dots, b_n; c; d_1 t, \dots, d_{n-1} t, d_n b_n^{-1} t).
\nonumber
\end{align}

Hence, using the Laplace transform of $\Phi_2^{(n)}(\cdot)$ in \cite[Eq. (9.4.55)]{srivastava1985multiple}, the Laplace transform of $\Phi_3^{(n)}(\cdot)$ is obtained as:
\begin{align}
&{\bf L}\left\{ \frac{t^{c-1}}{\Gamma(c)} \Phi_3^{(n)}(b_1, \dots, b_{n-1}; c; d_1 t, \dots, d_n t) \right\}
\nonumber\\
&\qquad
=   \lim_{b_n \to \infty}
    \frac{1}{s^{c}} \left( 1-\frac{d_n}{s} \frac{1}{b_n} \right)^{-b_n} \prod_{i=1}^{n-1} \left( 1-\frac{d_i}{s} \right)^{-b_i}
\nonumber\\
&\qquad
=   \frac{1}{s^{c}} e^{\frac{d_n}{s}} \prod_{i=1}^{n-1} \left( 1-\frac{d_i}{s} \right)^{-b_i}.
\label{eq:laplace_transf_Phi3}
\end{align}

Upon applying the above identity, we obtain \eqref{eq:cdf_snr}. This completes the proof of Theorem \ref{theo:theo_cdf}.
\end{IEEEproof}

\begin{table}[!h]
\renewcommand{\arraystretch}{1.7}
    \centering
    \caption{Relation between The fLoS Fading Model And Other Fading Models.}
    \begin{tabular}{c|c c c}
    \hline\hline
        Channels & \multicolumn{3}{c}{fLoS Fading Parameters}
    \\\hline\hline
        One-sided normal & $\underline{K} \to \infty$, &$\underline{k} = 1/2$, &$\underline{\lambda} = 0$
    \\\hline
        Nakagami-$m$ & $\underline{K} \to \infty$, &$\underline{k} = m$, &$\underline{\lambda} = 0$
    \\\hline
        Rayleigh & $\underline{K} = 0$, &$\forall\underline{k}$, &$\forall \underline{\lambda}$
    \\\hline
        Rice & $\underline{K} = K$, &$\underline{k} \to \infty$, &$\underline{\lambda} = 0$
    \\\hline
        Rician shadowed & $\underline{K} = K$, &$\underline{k} = m$, &$\underline{\lambda} = 0$
    \\\hline
        Hoyt 
        (Nakagami-$q$) & $\underline{K} = \frac{1-q^2}{2q^2}$, &$\underline{k} = 1/2$. &$\underline{\lambda} = 0$
    \\\hline
    \end{tabular}
    \label{tab:my_label}
\end{table}

The expressions in \eqref{eq:pdf_gamma}, \eqref{eq:cdf_snr}, and \eqref{eq:raw_moment} are novel, and their accuracy is validated through comprehensive Monte Carlo simulations. 
    However, the newly proposed fading model's expressions are quite challenging from a computational perspective as compared to the $\kappa$-$\mu$ fading case. 
In section \ref{sec:result}, we introduce an efficient method to numerically compute these functions.
    The relationship of the fLoS fading distribution with other well-known distribution is summarized in Table \ref{tab:my_label}, where $\Omega = \frac{1}{k+\lambda}$.

\subsection{Integer value of d.o.f. parameter, $k$}
When the parameter $k$ in the fLoS model is a positive integer, the PDF expression is simplified. Specifically, using the property of the $\Phi_3(\cdot)$ function in \cite[Eq. (8)]{Ermolova2014TSC}, we can derive a more tractable expression for the PDF, as detailed in the subsequent corollary.

\begin{Corollary}
\label{cor:pdf_simple}
Considering $\snr \sim {\cal F}_{\los}(\avgsnr; K; k, \lambda, \Omega)$ with non-negative real values $K$, $\lambda$, $\Omega$, and $k \in \mathbb{Z}^{+}$ is a positive integer, where $\avgsnr = \mean\{ \snr\}$ denotes the average SNR.
Then, the PDF of $\snr$ is:
%
\begin{align}
f_{\snr}(x) 
    &=   \frac{A^k}{\avgsnr \sigma^2}
    \exp\left( - A \frac{x}{\sigma^2 \avgsnr} - B\lambda \right) 
    \sum_{j=0}^{k-1}
        \frac{(-1)^j}{j!}  (1-k)_j
    \nonumber\\
    &\quad\times
        \left( \frac{B}{A\lambda} \frac{x}{\sigma^2 \avgsnr} \right)^{\frac{j}{2}}
        I_j\left( 2 \sqrt{ A B \lambda \frac{x}{\sigma^2 \avgsnr} } \right),~x > 0.
\label{eq:pdf_snr_int}
\end{align}
\end{Corollary}


We observe that the PDF is expressed as a finite mixture distribution, greatly simplifying the computational complexity of the fLoS fading distribution. Examining Corollary \ref{cor:pdf_simple} and utilizing the noncentral chi-square distribution definition in \eqref{eq:ncx2_pdf} and its relation to the $\kappa$-$\mu$ fading model, we observe the fLoS fading distribution as a finite mixture of $\kappa$-$\mu$ fading power distributions as:
\begin{align}
f_{\snr}(x) &= \tsum_{j=0}^{k-1} C_j 
    f_{\kappa\mu}(\underline{\avgsnr}; \underline{\kappa}, \underline{\mu}; x),~x > 0. \\
F_{\snr}(x) &= \tsum_{j=0}^{k-1} C_j 
    F_{\kappa\mu}(\underline{\avgsnr}; \underline{\kappa}, \underline{\mu}; x),~x > 0.
\end{align}
where  $\underline{\avgsnr} := \left( B\lambda + j + 1 \right) A \sigma^2 \avgsnr$,
    $\underline{\mu} := j + 1$,
    $\underline{\kappa} := \frac{B\lambda}{j+1}$,
    $C_j \triangleq \frac{(1-k)_j}{j!} (-B/A)^j A^k$ with $\sum_{j=0}^{k-1}{C_j} = 1$.
%
%
Here, $f_{\kappa\mu}(\underline{\avgsnr}; \underline{\kappa}, \underline{\mu}; x)$ and $F_{\kappa\mu}(\underline{\avgsnr}; \underline{\kappa}, \underline{\mu}; x)$ are the PDF and CDF of the $\kappa$-$\mu$ fading power as introduced in \cite[Eq. (2)]{Yacoub2007APM} and \cite[Eq. (3)]{Yacoub2007APM}, respectively.
%
Due to the finite mixture of $\kappa$-$\mu$ distributions, performance metrics for the fLoS fading channel (with integer $k$) can be readily derived from existing results for the $\kappa$-$\mu$ fading case, without resorting to further mathematical manipulations.
\section{Performance Analysis and Application}
\label{sec:performAna_App}
\subsection{Outage Probability}
The outage probability (OP), $P_{\rm out}$, indicates the probability that the instantaneous SNR drops below a set threshold, $\snr_{\rm th}$, given by $P_{\rm out} = \Pr\{ \snr < \snr_{\rm th} \}$ \cite{Bhardwaj2023GC}. 
    Consequently, the OP under fLoS fading is directly obtained by computing the CDF of the SNR, provided in \eqref{eq:cdf_snr}, as $P_{\rm out} = F_{\snr}(\snr_{\rm th})$.
In this context, when $\avgsnr \to \infty$ (high transmission SNR), we neglect $\Phi_3^{(3)}(\cdot)$ in \eqref{eq:cdf_snr} to obtain the asymptotic OP as:
\begin{align}
P_{\rm out}|_{\Uparrow}
    = A^k e^{-B\lambda} \snr_{\rm th} (\sigma^2 \avgsnr)^{-1}.
\label{eq:Pout_asymp}
\end{align}

A close observation of \eqref{eq:Pout_asymp} indicates that the fLoS fading model's diversity order, which is the rate at which the OP decreases in the high-SNR regime, is $d = 1$.


\subsection{Unified Ergodic Capacity}
The Ergodic capacity (EC) per unit bandwidth, in bps/Hz, is given by $\overline{C}   =  \frac{1}{\ln 2}\int_0^\infty \ln(1+\snr) f_{\snr}(\snr) {\rm d}\snr$ \cite{Yilmaz2012ICC}.
    However, the equation is difficult to derive due to the SNR PDF, $f_\snr(\snr)$, involving confluent hypergeometric functions.
To address this limitation, we propose a unified based method to approximate $\overline{C}$ [bps/Hz]. 
    Our method tailors the Prony's algorithm presented in \cite[Algorithm 2.1]{Zhang2019ACM} to approximate $f(x) = \ln(1+x)$~as:
\begin{align}
f(x) = 
    \tsum_{s=1}^{S} 
    \tsum_{k=1}^{M_s} { c_{s,k} e^{-T_{s,k} x} } \triangleq \hat{f}(x),~x \in [l_s, u_s].
\label{eq:ln_Prony}
\end{align}

The tailored Prony's approximation method is detailed in Algorithm \ref{algo:Prony}.
%
%
\begin{algorithm}
\label{algo:Prony}
  \caption{Tailored Prony's approximation method}
  \KwIn{$f$, $S$, $M_s$, $L_s \gg 2M_s$, $h_s$, $l_s$, and $u_s$, $s=1,\dots,S$, $u_{s+1}=l_s$;}
  \For{$s=1$ \KwTo $S$}{
      Compute $[{\bf f}]_i = f(x_i) \triangleq f_i$, $i = 1, \dots, L_s+1$ and $x_i = l_s + h_s(i-1)$ with $u_s = l_s+h_s L_s$;\\
      Perform Prony's method to approximate $f(x)$ for $x \in [u_s, l_s]$ and obtain the Prony's coefficients $T_k$ and $c_k$, $k = 1, \dots, M_s$;
  }
  \KwOut{Prony's coefficients;}
\end{algorithm}

From \eqref{eq:ln_Prony} and the fact that $\mean\{e^{s\snr}\} = \mgf_{\snr}(s)$, the EC per unit bandwidth is approximated as:
\begin{align}
\overline{C} 
    &= \frac{1}{\ln2} 
    \tsum_{s=1}^{S}
    \tsum_{k=1}^{M} c_{s,k}
    \Big[
        F(T_{s,k}, u_s) - F(T_{s,k}, l_s)
    \Big],
\label{eq:cap_approx}
\end{align}
where $F(T,x) = \int_0^{x} e^{-T s } f_{\snr}(s) {\rm d} s$, or \cite[Eq. (4)]{Lopez2017TC}
\begin{align}
F(T,x) = \mathbf{L}^{-1} \left\{ \frac{\mgf_{\snr}( -T - p )}{p} ; p, x \right\},
\end{align}
which is derived in a similar manner as $F_{\snr}(x)$ in 
\eqref{eq:cdf_snr}.

Observing that $\ln(1+x) \to \ln(x)$ as $x \to \infty$, we directly derive the asymptotic EC per unit bandwidth as:
\begin{align}
\overline{C}|_{\Uparrow}
    = \tsum_{s=1}^{S} \tsum_{k=1}^{M} \frac{c_{s,k} e^{T_{s,k}}}{\ln 2}
    \Big[
        F(T_{s,k}, u_s) - F(T_{s,k}, l_s)
    \Big].
\label{eq:cap_asymp}
\end{align}

\subsection{Multiple-input single-output (MISO) systems under channel aging}
The SNR distribution over fLoS fading channel finds application in many real-world scenarios.
Within these scenarios, the intended physical interpretation of the fLoS fading distribution may vary, yet the distribution itself remains useful in characterizing SNR distributions effectively. 

In downlink MISO systems, with a high-speed moving receiver Rx, channel aging poses a significant challenge, causing discrepancies between current and estimated channel state information (CSI) \cite{ChopraTWC2018}. 
    This mismatch degrades the wireless system performance and reliability. 
Specifically, the true CSI $\mathbf{h}$ conditioned on the estimated CSI $\hat{\mathbf{h}}$ is given by $ \mathbf{h} | \hat{\mathbf{h}} = \rho \widehat{\mathbf{h}}+ \bar{\rho} \mathbf{z}$, where $\mathbf{z} \sim {\cal CN}(\mathbf{0}_N, \mathbf{I}_N)$, with $N$~being the number of transmit antennas, $\rho$ being the temporal correlation coefficient, and $\bar{\rho} = \sqrt{1-\rho^2}$ \cite{ChopraTWC2018}. 
    The effective fading from transmitter to receiver, $h_{\eff}$, under Maximal Ratio Transmission (MRT) beamforming, is $h_{\eff}|\hat{\mathbf{h}} \sim {\cal CN}(\rho \Vert \hat{\mathbf{h}} \Vert, \bar{\rho}^2)$, where $\Vert \hat{\mathbf{h}} \Vert$ is the $L2$-norm of the estimated CSI.
Assuming Rician fading: $\hat{\mathbf{h}} \sim \mathcal{CN}\left( \frac{\sqrt{\kappa}}{\sqrt{\kappa+1}} \check{\bf h}, \frac{\mathbf{I}_N}{\kappa+1} \right) \in \mathbb{C}^{N}$, where $\kappa$ is the Rician factor and $\check{\bf h}$ represent the LoS components. Hence, the SNR at Rx is $\snr \triangleq \frac{P}{N_0 W} L_\eff |h_{\eff}|^2$, where $P$ [W] is the Rx's transmission power, $N_0$ [W/Hz] is the noise spectral density, $W$ [H] is the channel bandwidth, and $L_\eff$ denotes the effective path loss between Tx and Rx.
    Based on \eqref{eq:S_2} and noting that $\Vert \hat{\mathbf{h}} \Vert^2 \sim \widetilde{\chi}_{N}^2(\frac{1}{\kappa+1}, N\kappa)$, the true distribution of the received SNR, $\snr$, is given by:
\begin{align}
\snr \sim {\cal F}_{\los}\left( 
    \avgsnr; {\rho^2}/{\bar{\rho}^2}, N, \kappa N, [\kappa+1]^{-1}
\right),
\end{align}
where $\avgsnr \triangleq \frac{P L_\eff}{N_0 W}$. Thus, when Rx moves with high speed and causes channel aging, the effective channel's true diversity order between Tx and Rx remains $1$ regardless the number of transmit antennas, as discussed in the previous section.

\section{Numerical results}
\label{sec:result}

Although the confluent hypergeometric functions $\Phi_3^{(n)}(\cdot)$, ${\Phi_3(\cdot) = \Phi_3^{(2)}(\cdot)}$, and $\Psi_2(\cdot)$ are not available in MATLAB nor in Mathematica, we can numerically compute them as:
\begin{align}
&\Phi_3^{(n)}(b_1, \dots, b_{n-1}; c; d_1 t, \dots, d_n t)
\nonumber\\
    &\qquad= \frac{\Gamma(c)}{t^{c-1}}
    \int_{L-jT}^{L+jT}{ s^{-c} \frac{e^{st+\frac{d_n}{s}}}{j2\pi}
       \prod_{i=1}^{n-1} \left( 1-\frac{d_i}{s} \right)^{-b_i} {\rm d} s }, \label{eq:Phi3_matlab} \\
&\Psi_2(a; d; a; y, xt)
    = e^{y+xt} \Phi_3(d-a; d, a; -xt, x y t),
\end{align}
where $L = \max(d_1, \dots, d_{n-1}, 0)+\epsilon$ with $\epsilon>0$ and $T\gg 0$ are manually chosen to ensure the computation accuracy. For example, we set $\epsilon = \frac{1}{2}$ and $T = 10^3$ in all the simulations.

\subsection{Effect of fading parameters}
\begin{figure}[!h]
    \centering
    \includegraphics[width=0.8\linewidth]{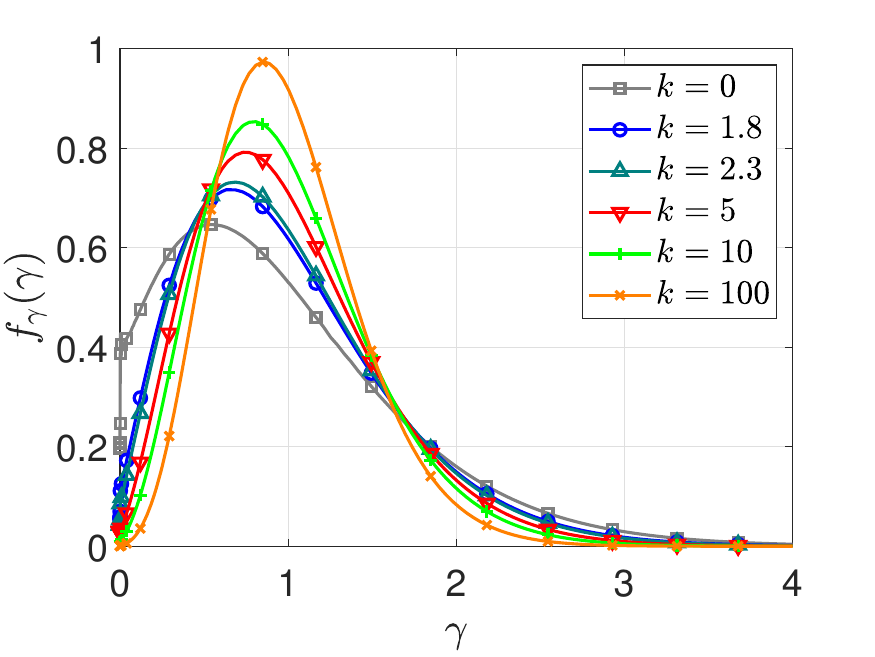}
    \caption{PDF of the SNR over the fLoS fading with different values for the d.o.f., $k$, where $\avgsnr = 0$ dB, $\lambda = 5$ and $K = 10$ dB.}
    \label{fig:pdf_vs_dof}
\end{figure}
\begin{figure}[!h]
    \centering
    \includegraphics[width=0.8\linewidth]{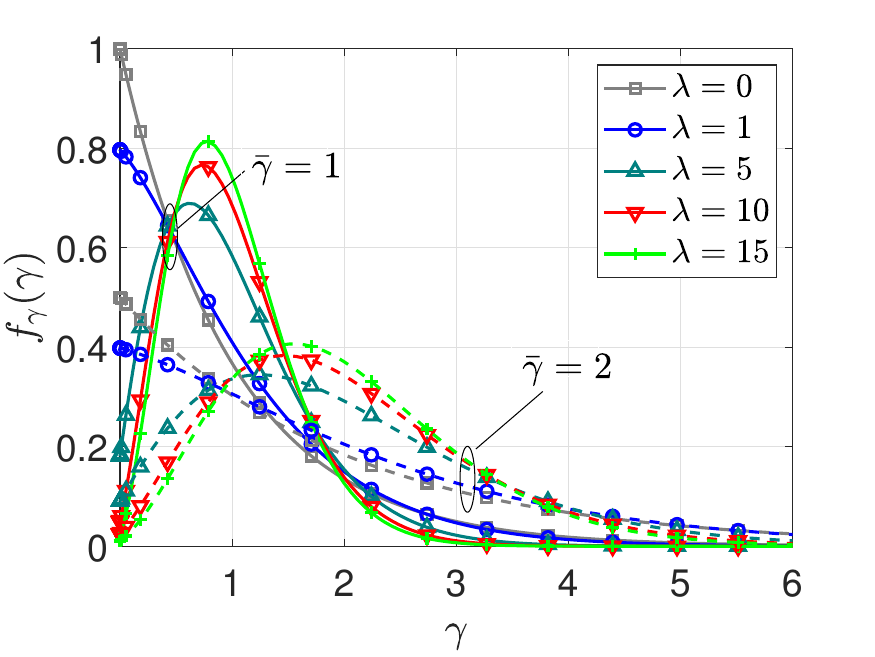}
    \caption{SNR PDF over fLoS fading versus $\lambda$ with $k = 1$ and $K = 10$ dB. Solid lines correspond to the $\avgsnr = 1$ case. Dashed lines correspond to the $\avgsnr = 2$ case.}
    \label{fig:pdf_vs_lambda}
\end{figure}

We now demonstrate how the fading parameters of the fLoS fading model influence the distribution shape. Due to space constraints, we concentrate on the SNR PDF and use $\Omega = \frac{1}{k+\lambda}$ to ensure $\mean\{\xi^2\} = 1$ and $\mean\{\snr\} = \avgsnr$.

Fig. \ref{fig:pdf_vs_dof} illustrates the changes in the PDF as the d.o.f. parameter varies. Increasing $k$ reduces the probability for having very low SNR values by concentrating the SNR around its mean value $\avgsnr$. Additionally, higher values of $k$ alleviate the fading severity, with $k = 0$ indicating the most severe fading conditions.

Fig. \ref{fig:pdf_vs_lambda} depicts the changes in the PDF as the noncentrality parameter $\lambda$ varies. Similarly, increasing $\lambda$ reduces the probability for very low SNR values by shifting the distribution to the right causing the peak to move away from zero, while the PDF is not concentrated around its mean as compared to increasing the d.o.f.. Smaller values of $\lambda$ also result in distributions with heavier tails, reducing the probability for having low SNRs. 

\subsection{Performance Analysis}
\begin{figure}[!h]
    \centering
    \includegraphics[width=0.8\linewidth]{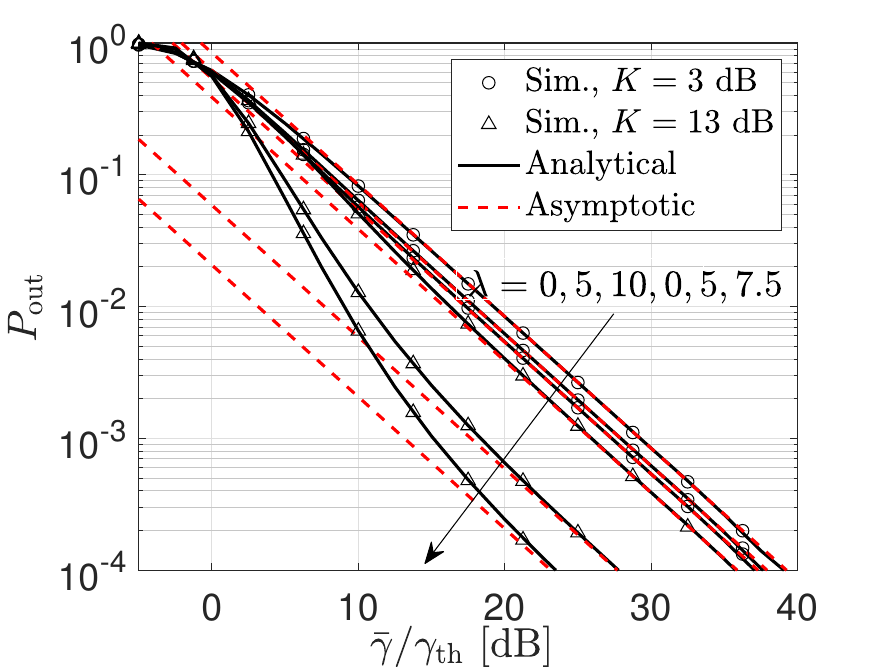}
    \caption{OP as a function of the normalized $\avgsnr$ for fLoS fading with $k = 1.5$, $K = 13$ dB and different values of $\lambda$.}
    \label{fig:op}
\end{figure}
Fig. \ref{fig:op} presents the OP as a function of the normalized average SNR, $\snr/\snr_{\rm th}$, with different instances of the noncentrality parameter, $\lambda$. Analytical and asymptotic results for the OP are obtained by applying \eqref{eq:cdf_snr} and \eqref{eq:Pout_asymp}, respectively. 
    Simulation results are presented validate the expressions under the current settings. 
We observe that the OP decreases as $\lambda$ is increased, especially with increased $K$, and that the asymptotic OP results are highly accurate for sufficiently large $\avgsnr/\snr_{\rm th}$.

\begin{figure}[!h]
    \centering
    \includegraphics[width=0.8\linewidth]{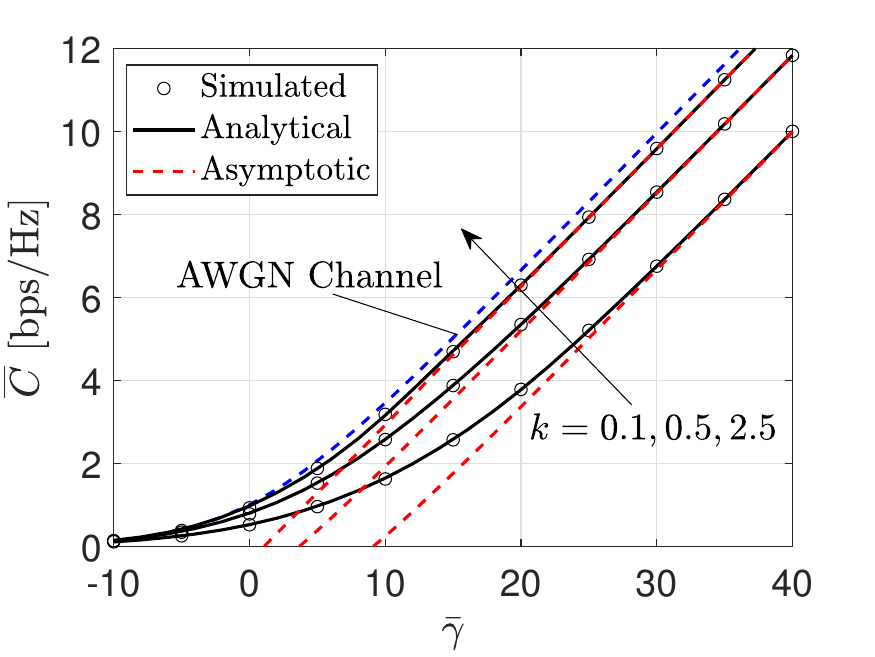}
    \caption{EC per unit bandwidth, $\overline{C}$, as a function of $\bar{\snr}$ for fLoS fading with $K = 13$ dB and $\lambda = 0$}
    \label{fig:ec}
\vspace{-10pt}
\end{figure}

Fig. \ref{fig:ec} depicts the EC considering various values for $k$. Simulation results (markers) verify the accuracy of \eqref{eq:cap_approx} and \eqref{eq:cap_asymp}. Moreover, the capacity of the AWGN case is included as a benchmark, as it always upper bounds the capacity of a fading channel due to Jensen’s inequality. As observed, the performance gap with respect to the AWGN case is reduced as $k$ is increased.

\subsection{Application}

\begin{figure}[!h]
    \centering
    \includegraphics[width=0.8\linewidth]{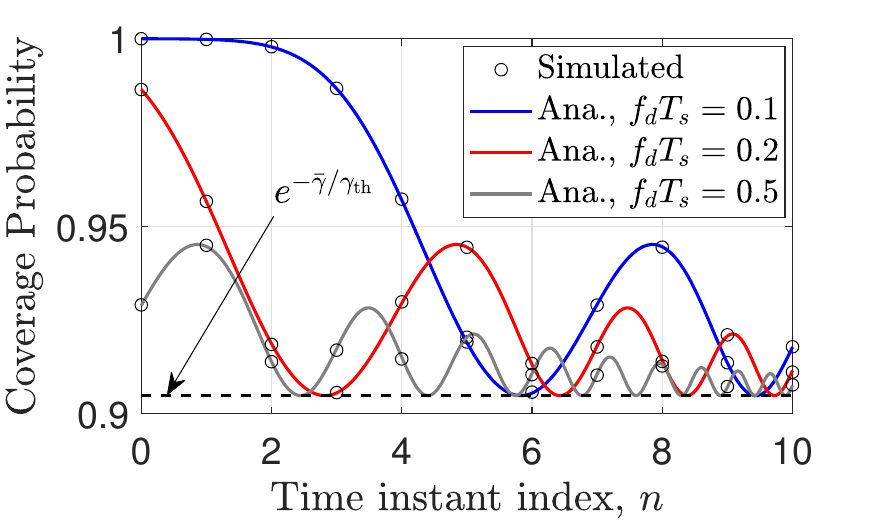}
    \caption{Coverage probability as a function of the time index, $n$, for fLoS fading with $N = 4$, $\kappa = 10$ dB, $\avgsnr / \snr_{\rm th} = 10$ dB, and different values of the normalized Doppler shift, $f_d T_s$.}
    \label{fig:covPrAging}
\end{figure}

Fig. \ref{fig:covPrAging} depicts the coverage probability, $P_{\rm cov} = 1 - P_{\rm out}$, versus the time instant index $n$ for varying normalized Doppler shifts, where the data transmission starts at $n = 0$.
    The downward trend illustrates a decreasing coverage probability with increasing $n$. Additionally, as $f_d T_s$ increases, the first zero position shifts leftward, indicating a reduction in the coherence time. 
The lower limit, $e^{-\avgsnr / \snr_{\rm th}}$, signifies an effective fading channel behavior similar to Rayleigh fading, one of the most unfavorable fading scenarios.

\section{Conclusion}
\label{sec:conclusion}

We introduced the new fLoS fading model, motivated primarily by its ability to encompass the lognormal shadowing and shadowed Rician fading models. This fLoS fading distribution simplifies into a finite combination of the more straightforward $\kappa$-$\mu$ distribution when $k$ is a positive integer. The performance analysis of wireless communication systems under the influence of the proposed fLoS fading model was examined using the OP and EC, where a unified Prony's method is proposed to approximate the EC. Finally, we presented an application for the fLoS fading distribution in characterizing wireless systems under the influence of channel aging.

\bibliographystyle{IEEEtran}
\bibliography{ref}

\end{document}